\documentclass[proceedings]{crckbked}

\usepackage{amsfonts}
\usepackage{amsmath}
\usepackage{graphicx}
\usepackage{courier}
\usepackage[symbol]{footmisc}

\renewcommand\arraystretch{1}%
\DeclareMathAlphabet{\mathsf}{OT1}{cmss}{bx}{n}%
\DeclareMathAlphabet{\mathit}{OT1}{cmr}{bx}{it}%
\def\func#1{\mathop{\rm #1}}%
\makeatletter%
\@ifundefined{mathletters}{%
  \newcounter{equationnumber}
  \def\mathletters{%
     \addtocounter{equation}{1}%
     \edef\@currentlabel{\theequation}%
     \setcounter{equationnumber}{\c@equation}%
     \setcounter{equation}{0}%
     \edef\theequation{\@currentlabel\noexpand\alph{equation}}%
  }%
}{}%
\makeatother%

\begin{document}

\begin{opening}

\title{Deep ocean influence on upper ocean \\ baroclinic instability saturation\thanks{To appear \emph{in} O. U. Velasco-Fuentes et al. (eds.), \textit{Nonlinear Processes in Geophysical Fluid Dynamics}, Kluwer Academic.}}

\author{M.~J. \surname{Olascoaga} and F.~J. \surname{Beron-Vera}}

\institute{RSMAS, University of Miami\\
4600 Rickenbacker Cswy.\\
Miami, FL 33149, USA}

\author{J. \surname{Sheinbaum}}

\institute{CICESE\\
Km. 107 Carretera Tijuana-Ensenada\\
22800 Ensenada, Baja Cfa., Mexico}

\runningauthor{M.~J. Olascoaga et al.}

\runningtitle{Deep ocean influence on baroclinic instability}

\begin{abstract}
In this paper we extend earlier results regarding the effects of
the lower layer of the ocean (below the thermocline) on the
baroclinic instability within the upper layer (above the
thermocline). We confront quasigeostrophic baroclinic instability
properties of a 2.5-layer model with those of a 3-layer model with
a very thick deep layer, which has been shown to predict spectral
instability for basic state parameters for which the 2.5-layer
model predicts nonlinear stability. We compute and compare maximum
normal-mode perturbation growth rates, as well as rigorous upper
bounds on the nonlinear growth of perturbations to unstable basic
states, paying particular attention to the region of basic state
parameters where the stability properties of the 2.5- and 3-layer
model differ substantially. We found that normal-mode perturbation
growth rates in the 3-layer model tend to maximize in this region.
We also found that the size of state space available for
eddy-amplitude growth tends to minimize in this same region.
Moreover, we found that for a large spread of parameter values in
this region the latter size reduces to only a small fraction of
the total enstrophy of the system, thereby allowing us to make
assessments of the significance of the instabilities.
\end{abstract}


\end{opening}

\renewcommand{\theequation}{\thesection.\arabic{equation}}

\section{Introduction}

Observations indicate that most of the world oceans variability is
confined in a thin layer limited from below by the permanent
thermocline. There, the density is approximately uniform in the
vertical but has important horizontal gradients. The latter imply
the existence of a considerable reservoir of potential energy
within this layer, stored in the isopycnals
tilt and available to feeding baroclinic instability processes \cite%
{Gill-Green-Simmons-74}. Haine and Marshall
(1998)\nocite{Haine-Marshall-98} have argued that these processes
are of outmost importance for the dynamics of the upper ocean.
These authors pointed out that baroclinic instability waves can be
efficient transport agents capable of stopping convective
processes, thereby exerting a large influence in the thermodynamic
state of the ocean.

Because the total depth of the ocean is much larger than that of the upper
thermocline layer, the reduced-gravity setting has been commonly adopted to
studying the upper ocean baroclinic instability \cite%
{Fukamachi-McCreary-Proehl-95,Ripa-JFM-95,Young-Chen-95,Beron-Ripa-97,
Ripa-DAO-99,Olascoaga-Ripa-99,Ripa-Taxco-99,Ripa-AMS-99,Ripa-JFM-00,Ripa-Barcelona-00,Ripa-JFM-01}%
. In this setting the active fluid layer is considered as floating on top of
a quiescent, infinitely deep layer. Olascoaga (2001)\nocite{Olascoaga-01b}
showed, however, that a thick---but finite---abyssal active layer can
substantially alter the stability properties of the upper ocean for certain
baroclinic zonal flows, such as the Atlantic North Equatorial Current (ANEC) %
\cite{Beron-Olascoaga-03}. Olascoaga (2001) considered spectral
(i.e. linear, normal-mode), formal (or Arnold), and nonlinear (or
Lyapunov) stability (Holm et
al., 1985; cf. also McIntyre and Shepherd, 1987)\nocite%
{Holm-Marsden-Ratiu-Weinstein-85,McIntyre-Shepherd-87} in a 3-layer
quasigeostrophic (QG) model. Primary attention was given to the limit of a
very thick bottom layer. The stability results were compared with those from
a reduced-gravity 2-layer (or 2.5-layer) model \cite{Olascoaga-Ripa-99}, and
assessments were made of the influence of the deep ocean on upper ocean
baroclinic instability.

To make further assessments, in this paper we turn our attention to
baroclinic instability saturation. Employing Shepherd's (1988) method\nocite%
{Shepherd-88b}, we establish and confront rigorous bounds on nonlinear
instability saturation in 2.5- and 3-layer models. This method, which builds
on the existence of a nonlinear stability theorem, has been previously used
to compute saturation bounds in 2.5- \cite{Olascoaga-Ripa-99} and 3-layer %
\cite{Paret-Vanneste-96} models. In addition to considering more
general model configurations than in these earlier works, we focus
on the size of state space available for the growth of eddies in
the region of parameter space where the models present
discrepancies in their stability properties. Also, unlike Paret
and Vanneste (1996), who computed numerical energy-norm bounds
based on both Arnold's first and second theorems, we derive
analytical expressions for enstrophy-norm bounds based on Arnold's
first theorem. Maximum normal-mode perturbation growth rates in
the 3-layer model are also calculated and contrasted with those in
the 2.5-layer model.

The reminder of the paper has the following organization. Section
2 presents the 3-layer model, from which the 2.5-layer model
follows as a limiting case. Normal-mode perturbation growth rates
are computed in \S\,3, along with an exposition of the main
results of formal and nonlinear stability analyses. Nonlinear
saturation bounds are then derived in \S\,4. We want to remark
that the number of basic state parameters that define a 3-layer
flow is too large to be explored in full detail. To facilitate the
analysis we reduce in some cases this number by fixing certain
parameters to values that can be taken as ``realistic,'' because
of being found appropriate for a region of the ocean mainly
dominated by the ANEC, which is a good example of a major zonal
current. Section 5 presents a discussion and the conclusions.
Appendices A and B are reserved for mathematical details relating
to the computation of the saturation bounds in the 3- and
2.5-layer models, respectively.

\setcounter{equation}{0}

\section{The Layer Models}

Let $\mathbf{x}$ denote the horizontal position with Cartesian coordinates $%
x $ (eastward) and $y$ (northward), let $t$ be the time, and let $D$ be an
infinite (or periodic) zonal channel domain on the $\beta $ plane with
coasts at $y=\pm \frac{1}{2}W.$ The unforced, inviscid evolution equations
for QG motions in a\textit{\ \textbf{3-layer model}}, with rigid surface and
flat bottom, are given by (cf. e.g. Ripa, 1992)\nocite{Ripa-JFM-92b}%
%TCIMACRO{
%\TeXButton{inicio letras}{\begin{mathletters}
%\label{QG}}}%
%BeginExpansion
\begin{mathletters}
\label{QG}%
%EndExpansion
\begin{equation}
\partial _{t}q_{i}=\mathbf{\hat{z}}\cdot \nabla q_{i}\times \nabla \psi
_{i},\quad \dot{\gamma}_{i}^{\pm }=0,
\end{equation}%
where $\psi _{i}$, being a nonlocal function of $\mathbf{q}:=(q_{i})^{%
\mathrm{T}}$ and $\boldsymbol{\gamma}:=(\gamma _{i}^{\pm })^{\mathrm{T}}$,
is uniquely determined by
\begin{equation}
\nabla ^{2}\psi _{i}-\sum_{j}\mathsf{R}_{ij}\psi _{j}=q_{i}-f
\end{equation}%
on $D,$ where%
\begin{equation}
\mathsf{R}:=\frac{1}{(1+r_{1})R^{2}}\left[
\begin{array}{ccc}
1 & -1 & 0 \\
-r_{1} & (1+s)r_{1} & -sr_{1} \\
0 & -s\frac{r_{1}r_{2}}{1+r_{1}} & s\frac{r_{1}r_{2}}{1+r_{1}}%
\end{array}%
\right] ,
\end{equation}%
and
\begin{equation}
\int \mathrm{d}x\,\partial _{y}\psi _{i}=\gamma _{i}^{\pm },\quad \partial
_{x}\psi _{i}=0
\end{equation}%
%TCIMACRO{\TeXButton{fin letras}{\end{mathletters}}}%
%BeginExpansion
\end{mathletters}%
%EndExpansion
at $y=\pm \frac{1}{2}W.$ Here, $q_{i}(\mathbf{x},t)$, $\psi _{i}(\mathbf{x}%
,t)$, and $\gamma _{i}^{\pm }=\mathrm{const.}$ denote the QG potential
vorticity, streamfunction, and Kelvin circulation along the boundaries of
the channel, respectively, in the top ($i=1$), middle ($i=2$) and bottom ($%
i=3$) layers. The Coriolis parameter is represented as $f=f_{0}+\beta y$,
the Nabla operator $\nabla =(\partial _{x},\partial _{y}),$ and $\mathbf{%
\hat{z}}$ denotes the vertical unit vector. The quantities
\begin{equation}
R^{2}:=\frac{g_{1}\bar{H}_{1}\bar{H}_{2}}{f_{0}^{2}\bar{H}},\quad s:=\frac{%
g_{1}}{g_{2}},\quad r_{1}:=\frac{\bar{H}_{1}}{\bar{H}_{2}},\quad r_{2}:=%
\frac{\bar{H}}{\bar{H}_{3}},
\end{equation}%
where $g_{i}$ is the buoyancy jump at the interface of the $i$-th and $(i+1)$%
-th layers, and $\bar{H}:=\bar{H}_{1}+\bar{H}_{2}$ with $\bar{H}_{i}$ the $i$%
-th layer reference thickness.

The\textit{\ \textbf{2.5-layer model} }follows from (\ref{QG}) in the limit
of infinitely thick ($r_{2}\rightarrow 0$) and quiescent ($\psi
_{3}\rightarrow 0$) lower layer. In the latter case, $%
(1+r_{1})(r_{1}/s)^{1/2}R$ and $R$ are equal to the first
(equivalent barotropic) and second (baroclinic) deformation
radius, respectively, in the limit of weak internal stratification
($s\rightarrow 0$).

The evolution of system (\ref{QG}) is constrained by the conservation of
\textit{\textbf{energy}}, \textit{\textbf{zonal}} \textit{\textbf{momentum}}%
, and an infinite number of vorticity-related \textit{\textbf{Casi\-mirs}},
which are given by%
\begin{equation}
\mathcal{E}:=-\tfrac{1}{2}\left\langle \psi _{i}q_{i}\right\rangle ,\quad
\mathcal{M}:=\left\langle yq_{i}\right\rangle ,\quad \mathcal{C}%
:=\left\langle C_{i}(q_{i})\right\rangle
\end{equation}%
(modulo Kelvin circulations along the boundaries), where $C_{i}(\cdot )$ is
an arbitrary function and $\langle \cdot \rangle :=$ $\sum_{i}\bar{H}%
_{i}\int\nolimits_{D}\mathrm{d}^{2}\mathbf{x}\,(\cdot ).$

\setcounter{equation}{0}

\section{Spectral, Formal, and Nonlinear Stability}

In this paper we deal with the stability of a \textbf{\textit{basic state}},%
\textbf{\ }i.e. equilibrium or steady solution of (\ref{QG}), of the form
\begin{equation}
\Psi _{i}=-U_{i}y,
\end{equation}%
which represents a baroclinic zonal flow. Here, $U_{i}=\sum_{i_1=i}^{2}g_{i_1}%
\sum_{i_2 =1}^{i_1}H_{i_2 ,y}/f_{0}$ $=\mathrm{const.}$, where
$H_{i}(y)$ is the thickness of the $i$-th layer in the basic
state, whereas $U_{3}$ is an arbitrary constant (set here to zero
with no loss of generality). The following six parameters are
enough to characterize the solutions of the
3-layer model stability problem:%
\begin{equation}
\kappa :=\sqrt{k^{2}+l^{2}}R,\;s,\;b:=\frac{\beta \bar{H}_{1}\bar{H}_{2}}{%
f_{0}\bar{H}H_{1,y}}\equiv \frac{\beta R^{2}}{U_{\mathrm{s}}},\;b_{\mathrm{T}}:=\frac{%
H_{,y}}{H_{1,y}}\equiv
\frac{sU_{2}}{U_{\mathrm{s}}},\;r_{1},\;r_{2}.
\end{equation}%
The first parameter, $\kappa $, is a nondimensional\textbf{\ \textit{%
wavenumber} }of the perturbation, where $k$ and $l$ are the zonal and
meridional wavenumbers, respectively. The second parameter, $s$, is a
nondimensional measure of the\textit{\ \textbf{stratification}.} The third
parameter, $b$, is a\textit{\ \textbf{planetary Charney number}}, namely the
ratio of the planetary $\beta $ effect and the topographic $\beta $ effect
due to the geostrophic slope of the upper interface. Here, $%
U_{\mathrm{s}}:=U_{1}-U_{2}$ is the velocity jump at the interface
(i.e. the current
vertical ``shear''). The fourth parameter, $b_{\mathrm{T}}$, is a \textbf{%
\textit{topographic Charney number}} given by the ratio of the topographic $%
\beta $ effects due to the geostrophic slopes of the lower and upper
interfaces. Finally, the fifth (resp., sixth), $r_{1}$ (resp., $r_{2}$),
parameter is the\textbf{\ \textit{aspect ratio} }of the upper to
intermediate (resp., upper-plus-intermediate to lower) reference layer
thicknesses. The problem considered by Olascoaga (2001) had $r_{1}=1.$ In
turn, the $2.5$-layer problem treated by Olascoaga and Ripa (1999), which
can be recovered upon making $r_{2}\rightarrow 0$ and $\psi _{3}\rightarrow
0,$ also had $r_{1}=1.$

Choosing a Casimir such that $\delta (\mathcal{E}+\mathcal{C}-\alpha
\mathcal{M})=0$ for any constant $\alpha \emph{,}$ the \textit{\textbf{%
pseudo energy--momentum}},
\begin{equation}
\mathcal{H}_{\alpha }[\delta \mathbf{q}]:=(\Delta -\delta )(\mathcal{E}+%
\mathcal{C}-\alpha \mathcal{M})=\mathcal{E}[\delta \boldsymbol{\psi }]+%
\tfrac{1}{2}\langle C_{i,QQ}\delta q_{i}^{2}\rangle ,
\end{equation}%
where $\boldsymbol{\psi}:=(\psi _{i})^{\mathrm{T}}$, is an exact
finite-amplitude invariant, quadratic in the\textbf{\ \textit{perturbation} }%
$\delta q_{i}(\mathbf{x},t)$ on the basic state potential vorticity $%
Q_{i}(y) $. Here, the symbols $\Delta $ and $\delta $ stand for total and
first variations of a functional, respectively, and $C_{i}(Q_{i})=\int
\mathrm{d}Q_{i}\,(\alpha -U_{i})Y(Q_{i})$ [$Y$ is the meridional coordinate
of an isoline of $Q_{i}$], where%
\begin{eqnarray}
Q_{1} &=&f_{0}+\left( b+\rho \right) yU_{\mathrm{s}}/R^{2}, \\
Q_{2} &=&f_{0}+\left[ b+\rho r_{1}\left( b_{\mathrm{T}}-1\right)
\right] yU_{\mathrm{s}}/R^{2},
\\
Q_{3} &=&f_{0}+\left( b-\rho ^{2}r_{1}r_{2}b_{\mathrm{T}}\right)
yU_{\mathrm{s}}/R^{2},
\end{eqnarray}%
with $\rho :=\left( 1+r_{1}\right) ^{-1}$. Arnold's (1965; 1966)\nocite%
{Arnold-65,Arnold-66}\ method for proving\textbf{\ \textit{formal stability}
}of $Q_{i}$ relies upon the sign-definiteness of $\mathcal{H}_{\alpha }.$
For evaluating the latter, it is useful to make the Fourier expansion $%
\delta \mathbf{q}=\sum_{k,l}\mathbf{\hat{q}}(t)\mathrm{e}^{\mathrm{i}%
kx}\sin ly,$ which implies $\mathcal{H}_{\alpha }=\tfrac{1}{2}\sum_{k,l}%
(\mathbf{\hat{q}}^{\ast})^{\mathrm{T}}\mathsf{H}_{\alpha
}\mathbf{\hat{q}}$ for certain matrix $\mathsf{H}_{\alpha
}(\kappa,s,b,b_{\mathrm{T}},r_{1},r_{2})$ (cf. Beron-Vera and
Olascoaga, 2003, \S\,2.2.1), so that the sign of
$\mathcal{H}_{\alpha }$ is determined from the inspection of the
elements of $\mathsf{H}_{\alpha }$ (cf. Mu et al., 1994; Paret and
Vanneste, 1996; Ripa, 2000a).

\begin{figure}[t]
\centerline{\includegraphics[width=12cm,clip=]{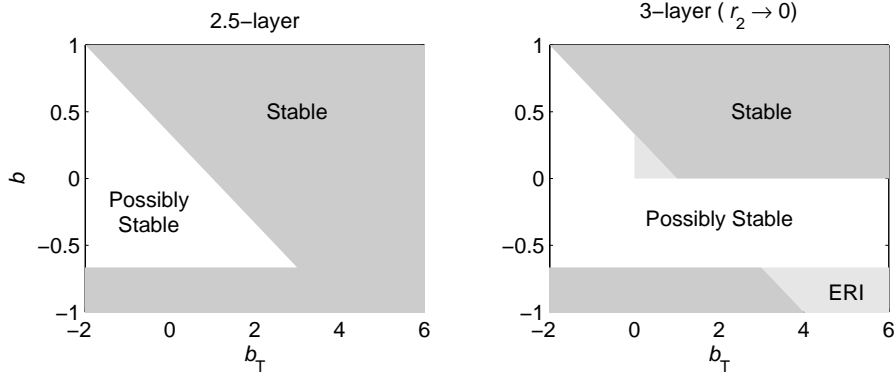}}
\caption{Stability/instability regions in the planetary $b$ vs.
topographic $b_{\mathrm{T}}$ Charney numbers space. Dark-shaded
regions are the locus of positive-definite pseudo energy--momentum
integrals. In the blank regions a pseudo energy--momentum integral
can be found to be negative definite if the zonal channel flow is
narrow enough. In the light-shaded regions no pseudo
energy--momentum integral can be proved to be sign definite.}
\label{arnold}
\end{figure}

In figure \ref{arnold} the regions of the
$(b,b_{\mathrm{T}})$-space labeled ``Stable'' correspond to basic
states for which there exists $\alpha $ such that
$\mathcal{H}_{\alpha }$ is positive definite (Arnold's first
theorem). The regions labeled ``Possibly Stable'' are locus of
basic states for which there exists $\alpha $ such that
$\mathcal{H}_{\alpha }$ is negative definite (Arnold's second
theorem) if the channel in which the flow is contained is
sufficiently narrow; cf. Mu (1998) and Mu and Wu (2001) for
details on optimality issues relating to Arnold's second theorem.
The results, which are independent of the choice of $s$, are
presented for $r_{1}=0.5$, a value estimated for the ANEC. The
r.h.s. panel in this figure corresponds to the 3-layer model in
the limit $r_{2}\rightarrow 0;$ the l.h.s. panel corresponds to
the 2.5-layer model. Clearly, as $r_{2}\rightarrow 0$ the 3-layer
model stable region does not reduce to that of the $2.5$-layer
model; it also requires $\delta \psi _{3}\rightarrow 0$
(Olascoaga, 2001). In the regions labeled ``ERI'' no sign-definite
$\mathcal{H}_{\alpha }$ can be found. Consequently, the
corresponding states are always unstable either through
normal-mode perturbations or explosive resonant interaction (ERI)
(Vanneste, 1995)\nocite{Vanneste-95b}. By contrast, in the
2.5-layer instability problem all basic states subject to ERI are
spectrally unstable. Finally,\textit{\ \textbf{nonlinear
stability}} can be proven for all formally stable states. Namely,
the departure from these basic states can be bounded at all times
by a multiple of the initial distance.

For\textit{\ \textbf{spectral stability} }a perturbation is assumed to be
infinitesimal and with the structure of a normal mode, i.e. $\mathbf{\hat{q}}%
=\varepsilon \mathbf{\tilde{q}}\mathrm{e}^{-\mathrm{i}kct}+O(\varepsilon ^{2})$%
, where $\varepsilon \rightarrow 0$. Nontrivial solutions for $\mathbf{\tilde{q%
}}$, which satisfies $\mathsf{H}_{c}\mathbf{\tilde{q}}=0,$ require
condition $\det \mathsf{H}_{c}=0$ to be fulfilled. This implies
the eigenvalue $c(\kappa ;s,b,b_{\mathrm{T}},r_{1},r_{2})$ to
satisfy $P(c)=0$, where $P(\cdot )$ is a cubic characteristic
polynomial (cf. Beron-Vera and Olascoaga, 2003, appendix B).

\begin{figure}[t]
\centerline{\includegraphics[width=12cm,clip=]{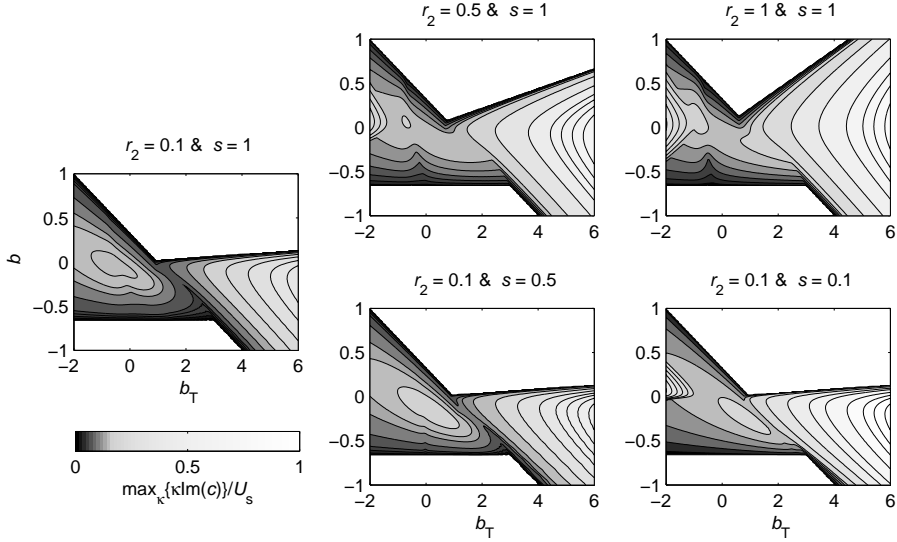}}
\caption{Maximum 3-layer model perturbation growth rate in the
planetary $b$ vs. topographic $b_{\mathrm{T}}$ Charney numbers
space for a fixed value of the aspect ratio $r_1$($=0.5$) of the
upper to intermediate reference layer thicknesses, and various
values of the aspect ratio $r_2$ of the upper-plus-intermediate to
lower reference layer thicknesses and the stratification parameter
$s$.} \label{maxtasa}
\end{figure}

Figure \ref{maxtasa} shows the 3-layer model maximum perturbation
growth rate, $\max_{\mathbf{\kappa }}\{\kappa \func{Im}c\},$ for
$r_{1}=0.5,$ and different values of parameters $r_{2}$ and $s$ in
the planetary $b$ vs. topographic $b_{\mathrm{T}}$ Charney numbers
space. In
general, the maximum perturbation growth rate increases with increasing $r_{2}$ and decreasing $s$%
. As $b_{\mathrm{T}}$ increases, the maximum perturbation growth
rate tends to achieve the largest values in the region where the
2.5-layer model is nonlinearly stable as a consequence of Arnold's
first theorem, even for (realistically) small values of $r_{2}$ as
depicted in the bottom panels of the figure.

\begin{figure}[t]
\centerline{\includegraphics[width=12cm,clip=]{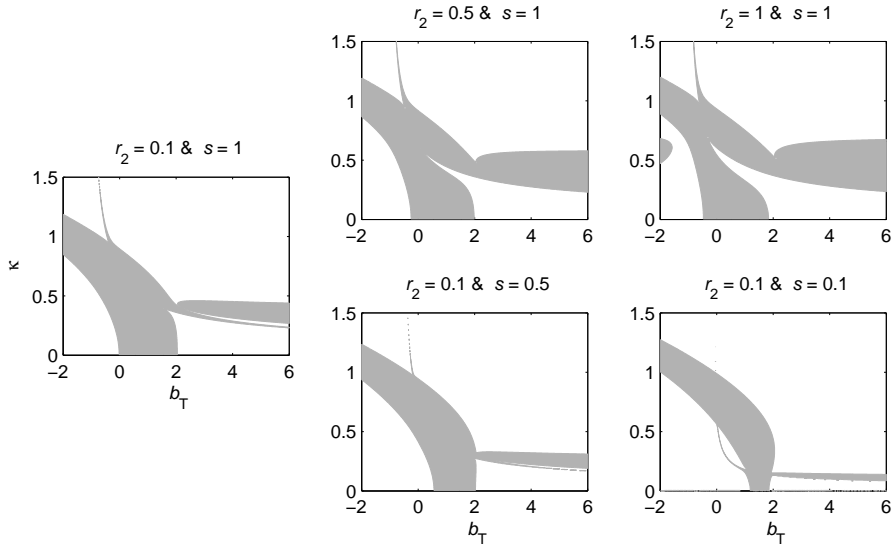}}
\caption{Three-layer model instability regions in the
nondimensional wavenumber $\kappa$ vs. topographic Charney number
$b_{\mathrm{T}}$ space, for a fixed value of the planetary Charney
number $b$($=-0.35$), and the same values of the aspect ratio
parameters $r_1$ and $r_2$, and the stratification parameter $s$
as in figure 2.} \label{deltas}
\end{figure}

Figure \ref{deltas} shows instability regions in $(\kappa ,b_{\mathrm{T}})$%
-space for $b=-0.35,$ a value estimated for the ANEC, and the same
values of parameters $r_{1},$ $r_{2},$ and $s$ as in figure
\ref{maxtasa}. The area of the region of possible wavenumbers for
destabilizing perturbations in the 3-layer model increases with
increasing $r_{2}$ and decreases with decreasing $s.$ Thus the
likelihood of these instabilities, which are not present in the
2.5-layer model, appear to be quite limited because they are
confined only to small bands of wavenumbers for small $s$ and
$r_2$. Yet the perturbation growth rates in these bands are not
negligible even for very small values of $r_{2}$ according to
Olascoaga (2001), who explained these instabilities as a result of
the resonant interaction between a neutral mode of the 2.5-layer
model instability problem and a short Rossby wave in the bottom
layer of the 3-layer model. In the next section we will see,
however, how the existence of nonlinearly stable states
contributes to arrest the eddy-amplitude growth, restricting the
significance of these instabilities, at least for certain basic
state parameters.

Let us finally turn our attention to the region of $(b,b_{\mathrm{T}})$%
-space where the two models allow for the possibility of
instability. The 2.5-layer model acquires its maximum perturbation
growth rates for $b=\frac{1}{2}\rho \lbrack (1-b_{\mathrm{T}})r_{1}-1]=-\frac{1}{6}%
(1+b_{\mathrm{T}})$ and $b_{\mathrm{T}}<1+r_{1}^{-1}=3$ as
$s\rightarrow 0,$ which corresponds to an exact cancellation of
the planetary and topographic $\beta $ effects, i.e.
$Q_{1,y}+Q_{2,y}=0$ (Olascoaga and Ripa, 1999). This result does
not hold for the 3-layer model because of the presence of
instabilities not present in the 2.5-layer problem. The latter
instabilities are confined to very narrow branches in the $(\kappa ,b_{%
\mathrm{T}})$-space and were also explained by Olascoaga (2001) as
a result of the resonant interplay of a neutral mode in the
2.5-layer model instability problem and a short Rossby wave in the
bottom layer of the 3-layer model. The maximum perturbation growth
rates associated with these instabilities are larger than those of
the 2.5-layer model (Olascoaga, 2001). In the following section we
will see, however, that the fraction of the total enstrophy of the
system available for eddy-amplitude growth can be much smaller in
the 3-layer model than in the 2.5-layer model for certain unstable
basic states.

\setcounter{equation}{0}

\section{Upper Bounds on Instability Saturation}

When a basic state is unstable, a priori upper bounds on the
finite-amplitude growth of the perturbation to this state can be
obtained using Shepherd's (1988) method. This method relies upon
the existence of a nonlinear stability theorem, and the bounds are
given in terms of the ``distance'' between the unstable basic
state, $Q_{i}^{\mathrm{U}}$ say, and the nonlinearly stable state,
$Q_{i}^{\mathrm{S}}$ say, in the infinite-dimensional phase space.

Let $\delta q_{i}^{\prime }(\mathbf{x},t)$ be that part of the perturbation
representing the ``waves'' or ``eddies,'' which result upon subtracting from
the perturbation its zonal (i.e. along-channel) average. Let $\mathbb{S}$
denote the space of all possible nonlinearly stable basic states, let $%
\left\| \mathbf{a}\right\| ^{2}:=\langle a_{i}^{2}\rangle /\mathcal{Z},$
where $\mathcal{Z}:=\langle (Q_{i}^{\mathrm{U}})^{2}\rangle ,$ and assume $%
q_{i}\approx Q_{i}^{\mathrm{U}}$ at $t=0$ so that $\mathcal{Z}$ corresponds to the%
\textit{\ \textbf{total enstrophy} }of the system. According to
Shepherd (1988), a rigorous\textit{\ \textbf{enstrophy-norm upper
bound on eddy-amplitude growth}}, based on Arnold's first theorem,
must
have the form%
\begin{equation}
\left\| \delta \mathbf{q}^{\prime }\right\| ^{2}\leq \frac{1}{\mathcal{Z}} \min_{Q_{i}^{\mathrm{S}}\in \,%
\mathbb{S}}\left\{ \frac{\max Q_{i,y}^{\mathrm{S}}}{Q_{i,y}^{\mathrm{S}}}%
\left\langle \left( Q_{i}^{\mathrm{U}}-Q_{i}^{\mathrm{S}}\right) ^{2}\right\rangle%
\right\}. \label{cota}
\end{equation}%
We want to mention that bounds---not treated here---on the zonal-mean perturbation, $\delta \bar{q}%
_{i}(y,t):=\delta q_{i}-\delta q_{i}^{\prime },$ or the total
perturbation
can also be derived as $\delta q_{i}$, $\delta \bar{q}%
_{i}$, and $\delta q_{i}^{\prime }$ satisfy the Pythagorean relationship $%
\left\| \delta \mathbf{q}\right\| ^{2}=\left\| \delta \mathbf{\bar{q}}%
\right\| ^{2}+\left\| \delta \mathbf{q}^{\prime }\right\| ^{2}$
(cf. Ripa, 1999c; Ripa, 2000b).

\begin{figure}[t]
\centerline{\includegraphics[width=12cm,clip=]{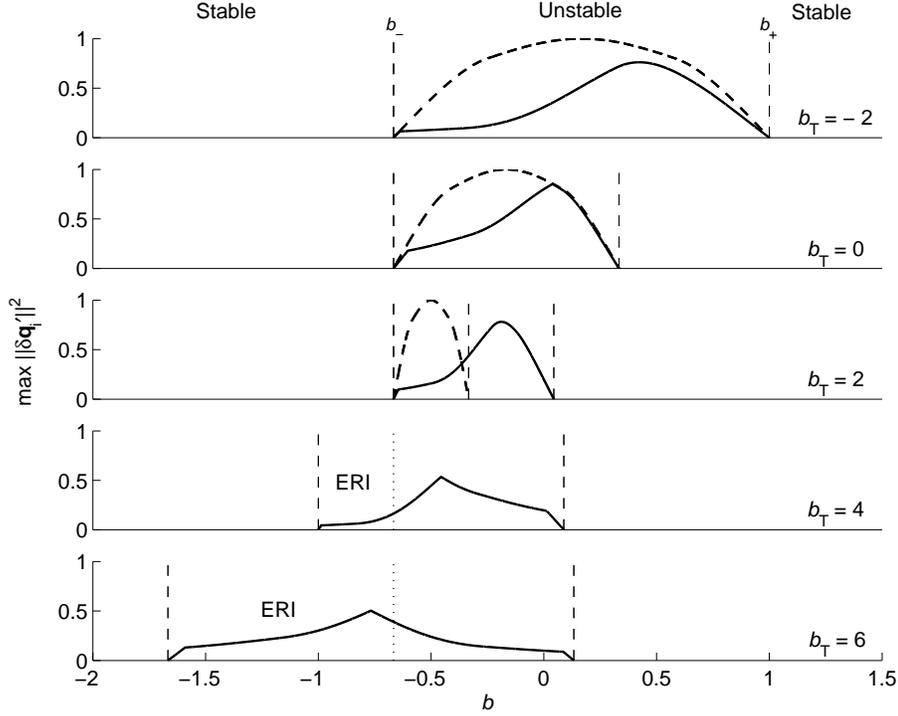}}
\caption{Fraction of the total potential enstrophy available for
eddy-amplitude growth in the 2.5- (dashed lines) and 3- (solid
lines) layer models as a function of the planetary Charney number
$b$, with aspect ratios $r_{1}=0.5$ and $r_{2}=0.1$ (the 2.5-layer
model has $r_{2}\rightarrow 0$), for different values of the
topographic Charney number $b_{\text{\textrm{T}}}$.} \label{cotas}
\end{figure}

Figure \ref{cotas} shows the tightest bound on instability
saturation corresponding to the 3-layer model (thick curves, cf.
appendix A) as a function of the planetary Charney number $b$, for
aspect ratios $r_{1}=0.5$
and $r_{2}=0.1,$ and various values of the topographic Charney number $b_{%
\mathrm{T}}$. The focus on the small value $r_{2}=0.1$ will allow
us to make comparisons with the stability properties of the
2.5-layer model. The latter model's bound (cf. appendix B) is also
plotted in the figure (dashed lines) assuming $r_{1}=0.5$ and the
same values of $b_{\mathrm{T}}$. It is important to remark that
the 2.5-layer bound does not follow from that of the 3-layer model
in the limit $r_{2}\rightarrow 0$; it also requires $\psi _{3}=0$
(Olascoaga, 2001). Both the 2.5- and 3-layer model bounds are
independent of the stratification parameter $s$. The 2.5-layer
model bound curves are only present in the upper three panels of
the figure because the 2.5-layer model predicts nonlinear
stability as a consequence of
Arnold's first theorem for $b_{\mathrm{T}}>1+r_{1}^{-1}=3$ and any value of $%
b$ (cf. also figure \ref{maxtasa}, lower-right panel).

Vertical dashed lines in each panel of figure \ref{cotas} indicate
the values of $b$ for marginal stability, denoted by $b_{\pm }$.
In the 3-layer model, $b_{-}=\min \{-\rho ,\rho r_{1}(1$
$-b_{\mathrm{T}})\}$ and $b_{+}=\max \{\rho
r_{1}(1-b_{\mathrm{T}}),\rho ^{2}r_{1}r_{2}b_{\mathrm{T}}\},$
whereas
for the 2.5-layer model, $b_{-}=-\rho $ and $b_{+}=\rho r_{1}(1-b_{\mathrm{T}%
}).$ Note that in the 2.5-layer model, while the $b_{-}$ marginal stability
value remains fixed at $b\approx -0.66667$, the $b_{+}$ moves toward smaller
values as $b_{\mathrm{T}}$ increases, until it collapses with $b_{-}$ at $b_{%
\mathrm{T}}=3$ (not shown in the figure). For $b>b_{+}$ and
$b<b_{-}$ the
basic flow in both the 2.5- and 3-layer models is nonlinearly stable. For $%
b_{+}<b<b_{-}$ the basic flow is unstable unless the zonal channel
flow is narrow enough for Arnold's second stability theorem to be
fulfilled. The latter is not always true in the 3-layer model
case, however, since there is a possibility that a spectrally
stable basic state could become unstable through ERI.

Three-layer surface-confined flows are susceptible to suffer more
destabilization than 2.5-layer flows. However, the state space
available, (determined by the fraction of total potential
enstrophy), for eddy-amplitude growth in the 3-layer model tends
to be smaller than the space available in the 2.5-layer model, at
least for certain basic state parameters. This is evident in the
upper three panels of figure \ref{cotas}. There is an overall tendency
of the 3-layer model bound to decrease as $b_{%
\mathrm{T}}$ increases. Moreover, for a large set of parameters
this bound reduces to only a small fraction of the total enstrophy
of the system. In these cases, the significance of the associated
instabilities is relative. On the other hand, there are basic
state parameters for which this fraction
is not negligible. As an example not shown in the figure, for $b=-0.35$ and $%
b_{\mathrm{T}}=2.5$, which are appropriate for a region similar to
the ANEC, the fraction of total enstrophy is about $45\%$, which
is not negligible. Of course, when the upper bounds are not small
enough, no unambiguous conclusion can be drawn about the
significance of an instability.

Figure \ref{cotas} also shows that the 3-layer model bound can be
significantly small for certain potentially ERI unstable flows
(cf. lower two panels in the figure). This also allows us to make
an unambiguous assessment of the significance of these type of
instabilities in the sense that they can be certainly negligible
for some basic state parameters.

Before closing this section, two points deserve additional
discussion. First, the result that the bounds for the 3-layer
model with a very thick deep layer are smaller than the 2.5-layer
model bounds in the region of parameters where the two models
share similar instability properties might seem at odds with the
fact that the 3-layer model is less constrained than the 2.5-layer
model, which allows for the development of more unstable states.
However, we believe that this result should not be surprising
inasmuch as the space over which the minimization is carried out
is larger in the 3-layer model than in the 2.5-layer model, which
offers the possibility of finding tighter bounds (cf. Olascoaga,
2001). Second, Paret and Vanneste (1996) were not able to draw a
conclusion on the significance of ERI instability as in the
present paper. These authors computed energy-norm saturation
bounds, according to both Arnold's first and second theorems,
using numerical minimization algorithms. These bounds, whose
analytical computation appears to be too difficult, were not found
to minimize at basic state parameters for which ERI instability is
possible. The analytical minimization involved in the derivation
of the enstrophy-norm of this paper has shown that the tightest
bounds are obtained using stable basic states whose parameters
have quite spread numerical values. The minimization thus requires
to search for a solution in a considerably large space, making
numerical computations extremely expensive. This might explain the
difficulty of Paret and Vanneste (1996) to find tighter bounds for
potentially ERI unstable basic flows.

\section{Concluding Remarks}

A previous study showed that the quasigeostrophic baroclinic
instability properties of a surface-confined zonal current may
differ substantially between a 2.5-layer model and a 3-layer, if
the former is considered to be a simplified 3-layer model with a
very thick deep layer. For certain basic state parameters, the
2.5-layer model predicts nonlinear stability whereas the 3-layer
model spectral instability. That study thus suggested that the
effects of the deep ocean on the baroclinic instability of the
upper thermocline layer of the ocean may be important for certain
currents.

In this paper we have made further assessments of the importance
of the deep ocean on upper baroclinic instability. We have
achieved this by analyzing (i) maximum normal-mode perturbation
growth rates and (ii) rigorous enstrophy-norm upper bounds on the
growth of perturbations to unstable basic states, in both 2.5- and
3-layer models of baroclinic instability. The new results show
that instabilities, which the 3-layer model predicts in the region
of basic state parameters where the 2.5-layer model predicts
nonlinear stability, appear to maximize their growth rates. At the
same time, however, the saturation bounds tend to minimize in this
same region of basic state parameters, thereby reducing the size
of state space available for eddy-amplitude growth. Moreover, for
a large subset of parameters in the region, the latter reduces to
only a small fraction of the total enstrophy of the system. In
these cases we have been able to make unambiguous assessments of
the significance of the associated instabilities in the sense that
they can be certainly negligible.

We close remarking that the important issue of making assessments
of the accuracy of the saturation bounds as predictors of
equilibrated eddy amplitudes is still largely open. This cannot be
addressed without performing direct numerical simulations. The
importance of this subject relies upon the potential of the bounds
in the architecture of transient-eddy parametrization schemes. The
treatment of these issues are reserved for future investigation.

\acknowledgements

We thank Ted Shepherd and an anonymous reviewer for helpful
comments. M.J.O. and F.J.B.V. were supported by NSF (USA). J.S.
was supported by CICESE's core funding and by CONACyT (Mexico).

\def\thesection{\Alph{section}}
\setcounter{section}{0}
\def\theequation{\Alph{section}.\arabic{equation}}
\setcounter{equation}{0}

\section{Three-Layer Model Bounds}

Upon minimizing the r.h.s. of (\ref{cota}) over all stable states,
we have been able to
find, in addition to the trivial bound $\max \left\| \delta \mathbf{q}%
^{\prime }\right\| ^{2}=1,$ various sets of possible bounds. A
first set involves 9 possibilities, for which
$Q_{I,y}^{\mathrm{S}}=\max
\{Q_{i,y}^{\mathrm{S}}\}$ and is given by%
%TCIMACRO{\TeXButton{llaves-ini}{\renewcommand{\arraystretch}{1.35}}}%
%BeginExpansion
\renewcommand{\arraystretch}{1.35}%
%EndExpansion
\begin{equation}
\max \left\| \delta \mathbf{q}^{\prime }\right\| ^{2}=\left\{
\begin{array}{l}
-Q_{i,y}^{\mathrm{U}}( Q_{I,y}^{\mathrm{U}}+Q_{i,y}^{\mathrm{U}})  \\
-\sum_{i}Q_{i,y}^{\mathrm{U}}\sum_{j}Q_{j,y}^{\mathrm{U}}%
\end{array}%
\right.
\end{equation}%
%TCIMACRO{\TeXButton{llaves-fin}{\renewcommand{\arraystretch}{1.00}}}%
%BeginExpansion
\renewcommand{\arraystretch}{1.00}%
%EndExpansion
$\div \frac{1}{4}\sum_{j}(Q_{j,y}^{\mathrm{U}})^{2},$ for $i\neq
I=1,2,3.$ A second set involves other 9 possibilities, for which
$\max
\{Q_{i,y}^{\mathrm{S}}\}=Q_{I_{1},y}^{\mathrm{S}}=Q_{I_{2},y}^{\mathrm{S}}$ and is given by%
%TCIMACRO{\TeXButton{llaves-ini}{\renewcommand{\arraystretch}{1.35}}}%
%BeginExpansion
\renewcommand{\arraystretch}{1.35}%
%EndExpansion
\begin{equation}
\max \left\| \delta \mathbf{q}^{\prime }\right\| ^{2}=\left\{
\begin{array}{l}
( Q_{I_{1},y}^{\mathrm{U}}) ^{2}+( Q_{I_{2},y}^{\mathrm{U}}) ^{2} \\
\frac{1}{2}( Q_{I_{1},y}^{\mathrm{U}}-Q_{I_{2},y}^{\mathrm{U}}) ^{2} \\
\frac{1}{2}( Q_{I_{1},y}^{\mathrm{U}}-Q_{I_{2},y}^{\mathrm{U}})
^{2}-2Q_{i,y}\sum_{j}Q_{j,y}^{\mathrm{U}}%
\end{array}%
\right.
\end{equation}%
%TCIMACRO{\TeXButton{llaves-fin}{\renewcommand{\arraystretch}{1.00}}}%
%BeginExpansion
\renewcommand{\arraystretch}{1.00}%
%EndExpansion
$\div \frac{1}{4}\sum_{j}(Q_{j,y}^{\mathrm{U}})^{2}$, for $i\neq I_{1},I_{2},$ where $%
\{I_{1},I_{2}\}=\{1,2\},\{2,3\},\{1,3\}.$ Another possibility
finally
results for $Q_{1,y}^{\mathrm{S}}=Q_{2,y}^{\mathrm{S}}=Q_{3,y}^{\mathrm{S}},$ and is given%
\begin{equation}
\max \left\| \delta \mathbf{q}^{\prime }\right\| ^{2}=\frac{2}{3}\left[ 1-%
\frac{Q_{1,y}^{\mathrm{U}}Q_{2,y}^{\mathrm{U}}+Q_{1,y}^{\mathrm{U}}Q_{3,y}^{\mathrm{U}}+Q_{2,y}^{\mathrm{U}}Q_{3,y}^{\mathrm{U}}}{%
\sum_{j}(Q_{j,y}^{\mathrm{U}})^{2}}\right] .
\end{equation}%
The tightest bound follows as the least continuous bound of the
above 20 possible bounds in the 4-dimensional space of unstable
basic state parameters, with coordinates
$(b,b_{\mathrm{T}},r_{1},r_{2})$ (the bounds are independent of
$s$).

\setcounter{equation}{0}

\section{Two-and-a-Half-Layer Model Bounds}

In the 2.5-layer model ($r_{2}\rightarrow 0$ and $\psi
_{3}\rightarrow 0$) the least bound in the 3-dimensional space of
unstable basic state
parameters, with coordinates $(b,b_{\mathrm{T}},r_{1}),$ is given by%
%TCIMACRO{\TeXButton{llaves-ini}{\renewcommand{\arraystretch}{1.35}}}%
%BeginExpansion
\renewcommand{\arraystretch}{1.35}%
%EndExpansion
\begin{equation}
\max \left\| \delta \mathbf{q}^{\prime }\right\| ^{2}=\left\{
\begin{array}{ll}
-4Q_{1,y}^{\mathrm{U}}(Q_{1,y}^{\mathrm{U}}+Q_{2,y}^{\mathrm{U}}) & \text{if }-\rho <b<b_{1} \\
\frac{1}{2}(Q_{2,y}^{\mathrm{U}}-Q_{1,y}^{\mathrm{U}})^{2} &
\text{if }b_{1}\leq b\leq b_{2}
\\
-4Q_{2,y}^{\mathrm{U}}(Q_{1,y}^{\mathrm{U}}+Q_{2,y}^{\mathrm{U}})
& \text{if }b_{2}<b<-r_{1}\rho ( b_{\mathrm{T}}-1) \tag{B.1}
\end{array}%
\right.
\end{equation}%
%TCIMACRO{\TeXButton{llaves-fin}{\renewcommand{\arraystretch}{1.00}}}%
%BeginExpansion
\renewcommand{\arraystretch}{1.00}%
%EndExpansion
$\div $ $[(Q_{1,y}^{\mathrm{U}})^{2}+(Q_{2,y}^{\mathrm{U}})^{2}],$ where $b_{1}:=-\frac{%
1}{4}\rho \left[ r_{1}(b_{\mathrm{T}}-1)+3\right] $ and $b_{2}:=-\frac{1}{4}%
\rho$ $\times\left[ 3r_{1}(b_{\mathrm{T}}-1)+1\right]$. This
result extends to arbitrary $r_{1}$ that of Olascoaga and Ripa
(1999).

\def\thesection{\arabic{section}}
\setcounter{section}{0}
\def\theequation{\arabic{section}.\arabic{equation}}
\setcounter{equation}{0}

\bibliographystyle{klunamed}

\end{document}